# Anomalous local distortion in BCC refractory high-entropy alloys

Yang Tong[1], Shijun Zhao[1], Hongbin Bei[1], Takeshi Egami[1,2,3], Yanwen Zhang[1,2], Fuxiang Zhang[1]

[1]*Division of Materials Science and Technology, Oak Ridge National Laboratory, Oak Ridge, TN 37831, USA.*

[2]*Department of Materials Science and Engineering, The University of Tennessee, Knoxville, TN 37996, USA.*

[3]*Department of Physics and Astronomy, The University of Tennessee, Knoxville, TN 37996, USA*

*This manuscript has been authored by UT-Battelle, LLC under Contract No. DE-AC05-00OR22725 with the U.S. Department of Energy. The United States Government retains and the publisher, by accepting the article for publication, acknowledges that the United States Government retains a non-exclusive, paid-up, irrevocable, worldwide license to publish or reproduce the published form of this manuscript, or allow others to do so, for United States Government purposes. The Department of Energy will provide public access to these results of federally sponsored research in accordance with the DOE Public Access Plan (http://energy.gov/downloads/doe-public-access-plan).*

**Abstract**

Whereas exceptional mechanical and radiation performances have been found in the emergent medium- and high-entropy alloys (MEAs and HEAs), the importance of their complex atomic environment, reflecting diversity in atomic size and chemistry, to defect transport has been largely unexplored at the atomic level. Here we adopt a local structure approach based on the atomic pair distribution function measurements in combination with density functional theory calculations to investigate a series of body-centered cubic (BCC) MEAs and HEAs. Our results demonstrate that all alloys exhibit local lattice distortions (LLD) to some extent, but an anomalous LLD, merging of the first and second atomic shells, occurs only in the Zr- and/or Hf-containing MEAs and HEAs. In addition, through the ab-initio simulations we show that charge transfer among the elements profoundly reduce the size mismatch effect. The observed competitive coexistence between LLD and charge transfer not only demonstrates the importance of the electronic effects on the local environments in MEAs and HEAs, but also provides new perspectives to in-depth understanding of the complicated defect transport in these alloys.

# 1. Introduction

High-entropy alloys (HEAs), or concentrated solid solution alloys (CSAs), are a new type of structural materials, with chemical complexity achieved by mixing of multiple elements (at least four) at equiatomic or near-equiatomic concentrations to form a single solid solution phase[1,2]. This new alloying strategy based on high configurational entropy greatly expands the composition space[3,4], offering opportunities for discovering alloys with new properties for advanced applications. CSAs based on refractory elements are of particular interest from both theoretical and practical points of view. With the yield strength at 1473 K far surpassing Ni-based superalloys[4,5], refractory CSAs (RCSAs) are potential high-temperature structural materials to meet the aggressive demand on raising operation temperature in next-generation jet engine and nuclear reactors, and re-entry vehicles. Recent experimental investigations on oxygen doping effect in a model TiZrHfNb RCSA have shown not only strength enhancement but also substantial ductility improvement[6], in contrast to the known interstitial-induced embrittlement in conventional refractory alloys. Moreover, some RCSAs exhibit abnormal behaviors, *e.g.*, anomalous phonon broadening[7], decomposition from one BCC phase to two BCC solid-solution phases[8-11]. All these phenomena hint at the existence of unexplored novel properties in RCSAs.

But deeper knowledge needs to be created for these chemically complex RCSAs to assist the alloy design without wandering in the endless compositional space. For this purpose, we focus on the local structure of RCSAs. Previously, strongly-distorted local structure was observed for the ZrNbHf refractory medium-entropy alloy[12]. In order to elucidate the physical origin of the strongly-distorted local structure in the RCSA, we examined the local structure of fifteen BCC equiatomic RCSAs with different combinations of refractory elements, by integrating pair distribution function measurement and density functional theory (DFT) calculations. The observed deviations of local structure from average structure demonstrate the existence of local lattice distortions (LLD) with varying extent in these alloys. The discovered charge transfer competes against the LLD to stabilize these metastable RCSAs. Our findings are expected to help design advanced structural materials towards excellent strength and radiation resistance by retarding defect transport.

## 2. Methods

**Sample preparation.** Elemental Ti, V, Zr, Nb, Mo, Hf, Ta, W and Re (>99% pure) were carefully weighted and mixed into fifteen equiatomic concentrated solid solution alloys by arc melting. The arc-melted buttons were flipped and re-melted at least five times to improve homogeneity. Then, fine powders were ground from the as-cast buttons for synchrotron X-ray characterization.

**Materials characterization.** The diffraction and total scattering measurement were carried out at 28-ID-2 beamline of the NSLS II with an X-ray energy of 67 keV ($\lambda$ = 0.185 Å) and 11-ID-B beamline of the APS with an X-ray energy of 58.66 keV ($\lambda$ = 0.2114 Å). A two-dimensional stationary detector with 200×200 $\mu m^2$ pixel size was used to collect data. Calibration was performed using either $CeO_2$ or Ni NIST powder standard. Fit2D software[24] was used to correct for a beam polarization and a dark current. X-ray diffraction profiles were analyzed with the Rietveld refinement using the program $GSAS^{25}$. PDFgetX2[26] was used to obtain real-space PDF by a Fourier transformation of the measured reciprocal-space structure function, $S(Q)$, in a $Q$ range of 25 Å$^{-1}$,

$$G(r) = \frac{2}{\pi} \int Q[S(Q) - 1] \sin(Qr) \, dQ$$

where $Q$ is the magnitude of scattering vector[27]. By using PDFGUI software[28] the measured PDFs were refined with different structure models.

**Density functional theory calculations.** DFT calculations were performed using the Vienna *ab initio* simulation package (VASP)[29]. The exchange and correlation interactions were described by a gradient corrected functional in the Perdew-Burke-Ernzerhof (PBE) form[30]. Standard projector-augmented-wave (PAW) pseudopotentials distributed with VASP were used to treat electron-ion interactions[31]. The energy cutoff for the plane-wave basis set was set to be 300 eV. A Γ-centered 2×2×2 k-points was used to sample the Brillouin zone. The energy and force convergence criterion were set to be $10^{-4}$ eV and 0.01 eV/Å. The chemical disorder of alloys was modeled by special quasi-random structures (SQS)[32] that were constructed using a simulated annealing Monte-Carlo algorithm. The supercells built based on the SQS contain 250 atoms. The optimal volume of the supercell is determined by calculating the energy-volume curve and then fixed during atomic relaxation.

3. Results

3.1 **Average structure by X-ray diffraction**

We determined the average structure of the RCSAs by synchrotron X-ray diffraction with a two-dimensional (2-D) detector. Two representative 2-D diffraction images are shown in Fig. 1a, b. The homogeneous diffraction intensity distribution of each diffraction ring in the 2-D diffraction images demonstrates the sample is powder without texture. However, strong intensity decay in the Bragg peaks at high diffraction angles was observed especially in the TiZrNbHfTa, TiZrNbHf and TiVZrNb RCSAs when compared to the seven-element

VNbTaTiMoWRe RCSA, indicating large local atomic displacements in these three RCSAs.

The 2-D diffraction images were further integrated over azimuthal angles to obtain the one-dimensional (1-D) powder diffraction profiles, as shown in Fig. 1c. The diffraction profiles of all fifteen RCSAs can be indexed to a pure BCC phase. Structural information including lattice constants and Debye Waller factor (DWF, $U_{iso}$) were extracted from these diffraction patterns based on the Rietveld method[13], as listed in Table 1. The TiZrNbHfTa, TiZrNbHf and TiVZrNb RCSAs have larger lattice constants than other RCSA without Zr and/or Hf since a large space is needed to accommodate the large Zr and/or Hf atoms[14]. The DWFs of the VNbTaTiMoWRe, VNbTaMoWRe and NbTaMoWRe RCSAs are comparable with the DWFs of their constituent metals, whereas the rest of RCSAs have DWFs one order of magnitude larger than the pure metal cases for which the phonon thermal factor is dominant[15]. Note that the TiZrNbHfTa, TiZrNbHf and TiVZrNb RCSAs have extraordinarily large DWFs when compared to the rest alloys. These abnormally-large DWFs indicate that the existence of large LLD in these three Zr- and/or Hf-containing RCSAs (hereafter denoted as ZH-RCSAs). But a quantitative analysis remains difficult from the average structure analysis because the DWF contains contributions from both dynamic and static atomic displacements.

### 3.2 Local structure by the PDF

We further conducted the pair distribution function (PDF) study to examine the local structure of the RCSAs. The PDF is obtained from a Fourier transformation of the total scattering data, i.e. Bragg peaks and diffuse scattering induced by local distortion, and describes the local structure of the RCSAs in terms of interatomic distance, as presented in

Fig. 2a. The RCSAs without Zr and/or Hf elements (hereafter denoted as X-RCSAs) have well-separated first and second PDF peaks corresponding to the first and second atomic shells. For the ZH-RCSAs, however, an unusual feature can be readily identified in their PDFs, in which the first and second PDF shells overlap to form a broad peak with a hump on its right shoulder (Fig. 2b) accompanied by a rapid damping of structure features with increasing $r$. The overlap of the first two PDF peaks in these ZH-RCSAs is a clear feature of a strong LLD because for phonon vibrations the first PDF peak is sharper due to an effect of correlated motion among neighboring atoms[16]. The LLD is also evidenced by the shift of the first PDF peak. As compared in Fig. 2b, the value of the first PDF peak position ($P_{1st}$) is larger than the value expected from the average structure ($P_{avg} = \frac{\sqrt{3}}{2} * a$, where $a$ is the lattice constant determined by the Bragg peaks), indicating the presence of internal strain in the first atomic shell of these ZH-RCSAs. This local lattice strain can be quantified as $\varepsilon_{1st} = (P_{1st} - P_{avg})/P_{avg}$, and the value of $\varepsilon_{1st}$ was determined for all 15 RCSAs (see Table 1). Five of these RCSAs have $\varepsilon_{1st}$ higher than the value reported so far for the face-centered cubic FeCoNiCrPd HEA (0.79%)[17,18].

The distortion in the second atomic shell of these ZH-RCSAs is not obvious from Fig. 2b. However, an anomalous shift of the second PDF peak exists in the ZH-RCSAs. To demonstrate this point, we fitted the undistorted BCC random-alloy structure to the measured PDFs and two representative fitting cases are shown in Fig. 3. Note that in this fitting process, the lattice constant was fixed as the value determined from the diffraction pattern. The difference curves in Fig. 3a, c reveal that the undistorted BCC model can be fitted to the average structure in the high $r$ range but not the local structure in the low $r$ region. However, this local structure deviation only happens within the first atomic shell for the VNbTaTa RCSA, whereas the LLD in the TiZrNbHf RCSA involves atoms in both the first and second atomic shells. After examining the fitting for all fifteen RCSAs, we found a general trend that the LLD in the X-RCSAs is localized within the first atomic shell while the LLD in the ZH-RCSAs extends up to the second atomic shell. In contrast to the expansion of the first PDF peaks, the second PDF peaks in these ZH-RCSAs surprisingly shift to the low $r$ region to reduce interatomic distances. Clearly the first and second atomic shells tend to merge in these ZH-RCSAs to form a combined atomic shell involving fourteen neighbors.

**3.3 Distorted BCC crystal structure by DFT simulation**

To elucidate the microscopic origin of the local distortions in the measured PDF through the DFT calculations, we firstly built supercell models with 250 atoms based on special quasi-random structure method, and then performed volume and ionic relaxations to find atomic positions for the lowest energy state (see Methods). Two representative relaxed supercells for the ZH-RCSAs and the X-RCSAs, respectively, are shown in Fig. 4a, c. The DFT results clearly demonstrate the major differences between ZH-RCSAs and X-RCSAs in the local structure. It can be seen that atoms in the TiZrNbHf RCSA displace much more severely from their perfect lattice sites when compared with the VNbTaTi RCSA. We then fitted the distorted structure models to the measured PDFs (Fig. 4b, d). Compared with the fitting based on undistorted structure (Fig. 3), the distorted structure model much more successfully reproduces both the average structure in the high-$r$ region and the local structure in the low-$r$ range, leading to a significant improvement on the goodness of the fit, Rw. In the fitting process, the lattice constant parameter was refined, and its value perfectly agrees with the one determined from the average structure.

To further understand the anomalous shift in the PDF peaks of the ZH-RCSAs, we examined the variations in the atomic pair distances particularly in the first and second atomic shells after the relaxation processes. We show the distributions of atomic pair distances for the 1NN and 2NN atoms (defined from the unrelaxed supercells) (Fig. 4e, f, g). The distributions of the 1NN and 2NN atomic distances are broad enough to overlap with each other, resulting in an undistinguished gap between the first and second atomic shells in the relaxed structure. The small atoms, Ti, Nb, Ta and V in these RCSAs, however, has the widest spreading of interatomic distances, as revealed from the distance distribution of the partial atomic pairs. In contrast, the distribution of the interatomic distances is relatively narrow for large Zr and Hf atoms, but these atoms tend to escape from the first and second atomic shells and meanwhile prefer staying in between the first and second atomic shells, causing the shift in the first and second PDF peak positions. The physical origin of these anomalous shifts is more complex, as illustrated later.

**3.4 Charge transfer effect**

The energy associated with LLD is exceptionally large[19], which destabilizes these ZH-RCSAs especially at intermediate temperatures where entropy contribution to the phase stability is decreased. For instance, the separation of large Zr and Hf atoms from other small ones was found in the TiZrNbHfTa RCSA through a decomposition of the single BCC phase to two NbTa- and ZrHf-rich BCC ones[8-11]. However, DFT calculations revealed that the Gibbs free energy of the TiZrNbHfTa and TiZrNbHf RCSAs rather decreases after relaxing atomic positions due to a considerable devaluation of enthalpy[19]. Therefore, besides the entropy effect, other factor related to chemistry competes with the substantial energy increase induced by LLD. In the fusion of different metals to form substitutional alloy, the difference in the Fermi levels among pure metals drives the displacement of electrons from one element to the other[20,21]. This charge transfer effect (CTE) not only changes the energy balance but also alters the atomic radius of each constituent element. To demonstrate this atomic size variation, we calculated the charge difference of the $d$ ($e_g$ and $t_{2g}$) orbitals for each atom in the supercells between the unrelaxed and relaxed states. Meanwhile, the relation of the charge difference in both $e_g$ and $t_{2g}$ orbitals with Wigner-Seitz (WS) cell volume is examined for some typical RCSAs (Fig. 5). Among all studied RCSAs, the charge transfer in the $t_{2g}$ orbitals follows a linear correlation with the WS volumes whereas charge transfer is not found in the $e_g$ orbitals. The distinct feature of charge transfer in different $d$ orbitals is related to the directional nature of the $e_g$ and $t_{2g}$ orbitals in BCC structure that the $t_{2g}$ orbital points to the 1$^{st}$ nearest neighbor, whereas the $e_g$ orbital points to the 2$^{nd}$ neighbors[22]. The results, presented in Fig. 5, reveal that in the RCSAs CTE tends to mitigate the size mismatch among their constituents by transferring electrons from the early transition elements (TEs) which are large to the later TEs which are smaller. We further calculated the average WS radius, $r_{ws}$, for each element in some exemplary RCSAs and compared them with the WS radius of pure metals in Table 2. The WS radii of these pure refractory metals are calculated from their lattice constant according to $3/4\pi r_{ws}^3 = a^3/N$, where $a$ and $N$ are lattice constant and number of atoms in a unit cell, respectively. We found that in these RCSAs the largest atoms, Zr and Hf, reduce their $r_{ws}$ by 4-6%, the $r_{ws}$ of the smallest V atom can increase it by 4-6.7%, and the $r_{ws}$ change in the medium-sized Ti, Nb and Ta atoms can be zero, positive or negative depending on the difference between their radii and the average radius of all elements in one RCSA, which are consistent with the findings in Fig. 5. The trend of transferring charge from large early

TEs to small ones qualitatively agrees with the electronegativity difference between two elements. For instance, the large Zr and Hf atoms are more electronegative than other small atoms.

In general a parameter based on the fixed metallic radii is widely used to estimate the size mismatch in HEAs, $\delta = \sqrt{\sum_{i=1}^{N} c_i \left(1 - r_i / \sum_{j=1}^{N} c_j r_j\right)^2}$, where $N$ is the total number of the constituent elements, $c_{i,j}$ and $r_{i,j}$ denote the atomic fraction and atomic radius of the $i$th or $j$th element, respectively[23]. The metallic radii of Zr and Hf elements are 9.4%-21.6% larger than the Ti, V, Nb and Ta elements[14], and the $\delta$ parameter estimates the size mismatch in the TiZrNbHfTa, TiZrNbHf, and TiVZrNb RSCAs to be 7.4%, 7.3% and 10.1%, respectively. Note that the $\delta$ value calculated from the WS radii of pure metals in Table 2 is equivalent to the one from the metallic radii since the ratio of WS radius to metallic radius for the BCC lattice is constant. However, after considering the CTE, the value of $\delta$ becomes to 1.2%, 1.4% and 1.8% for the TiZrNbHfTa, TiZrNbHf, and TiVZrNb RCSAs, respectively, which are remarkably smaller (~80% drop) than the estimations based on the metallic radii or WS radii of pure refractory metals. Therefore, we conclude that CTE can adjust the atomic radius of the constituents in the BCC RCSAs to profoundly relax their size mismatch and stabilize the solid solution.

## 4. Discussion and conclusion

The DFT calculation showed that the atomic mismatch in RCSAs considered here is greatly reduced because of the charge transfer from the early TEs and $4d$ elements which are large to later TEs and $3d$ elements which are smaller. Therefore one would expect reduced local distortion in all RCSAs. Then why the local environment in ZH-RCSAs is more distorted than in X-RCSAs? The answer is that the atomic size mismatch is reduced only in *average*, but the actual interatomic distances depends on chemistry. As shown in Fig. 5 the A-A distances are considerably smaller and the B-B distances are considerably larger than the average, even though the average size mismatch is small, because charge transfer is much weaker for the A-A pairs and B-B pairs whereas it is strong for the A-B pairs. Thus charge transfer increases the dispersion in the interatomic distances increases. In particular Zr and Hf are large, and have high Fermi levels. Thus the interatomic distances around Zr and Hf are widely distributed as shown in Fig. 5, resulting in the merger of the $1^{st}$ and $2^{nd}$ neighbor shells.

In summary, through pair distribution function technique, we investigated the local structure variation in fifteen loosely-packed BCC CSAs, obtained by mixing multiple refractory elements with different size and chemistry at the same crystallographic sites. We found these RCSAs have distorted local structure, characterized by the deviation of their local structure from average structure. However, the length-scales of this deviation are different between the ZH-RCSAs and the X-RCSAs and closely related to the large size and Fermi level differences between Zr/Hf atoms and other constituent atoms. For the X-RCSAs, the LLD only occurs within the first atomic shell, whereas the LLD in the ZH-RCSAs is larger as a consequence of more extensive charge transfer, involving atoms in both the first and second atomic shells. In particular, we revealed that whereas charge transfer greatly reduces the atomic size mismatch in average and contributes to the stability of the solid solution, it broadens the distribution of local interatomic distances. As a final remark, the obtained insights on not only local lattice distortion but also charge transfer effect in the BCC refractory concentrated solid solution alloys are fundamentally important for defect (dislocation, radiation-induced defects) transport physics and therefore have significant practical implications as they offer a quantitative guidance to design strong and radiation-tolerant structural materials.

**Acknowledgements**

This work was supported as part of the Energy Dissipation to Defect Evolution (EDDE), an Energy Frontier Research Center funded by the U.S. Department of Energy, Office of Science, Basic Energy of Sciences under contract number DE-AC05-00OR22725. The X-ray diffraction and total scattering measurement were conducted at the National Synchrotron Light Source II (NSLS II) and Advanced Photon Sources (APS). This

**Author contributions**

Y. T., H. B., and F. Z. designed the research; H. B. developed the alloy compositions; Y. T. and F. Z. conducted the synchrotron X-ray experiments; S. Z. performed the DFT calculations; Y. T., S. Z. T. E. and F. Z. analyzed the data. Y. T., T. E. and F. Z. wrote the manuscript. All authors discussed the results and commented on the manuscript.

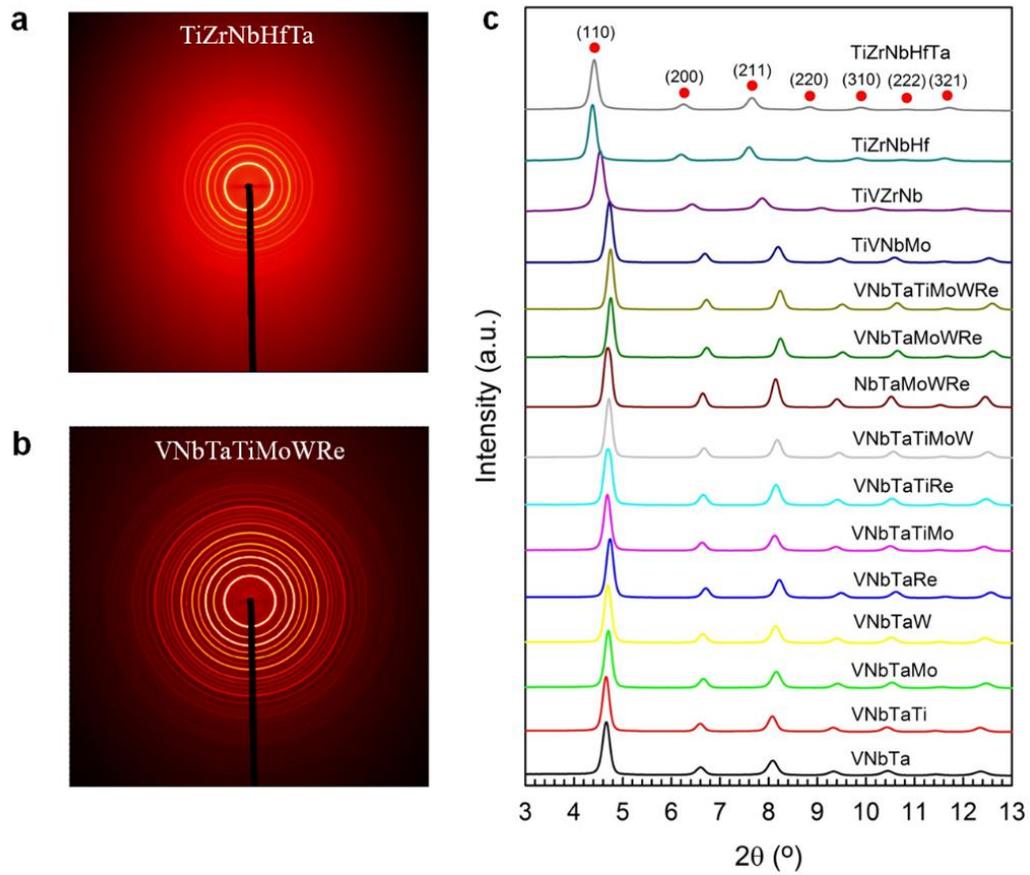

**Figure 1. Average structure of the RCSAs.** 2-D diffraction images of the TiZrNbHfTa (a) and VNbTaTiMoWRe (b) RCSAs. (c) 1-D diffraction profiles of fifteen RCSAs.

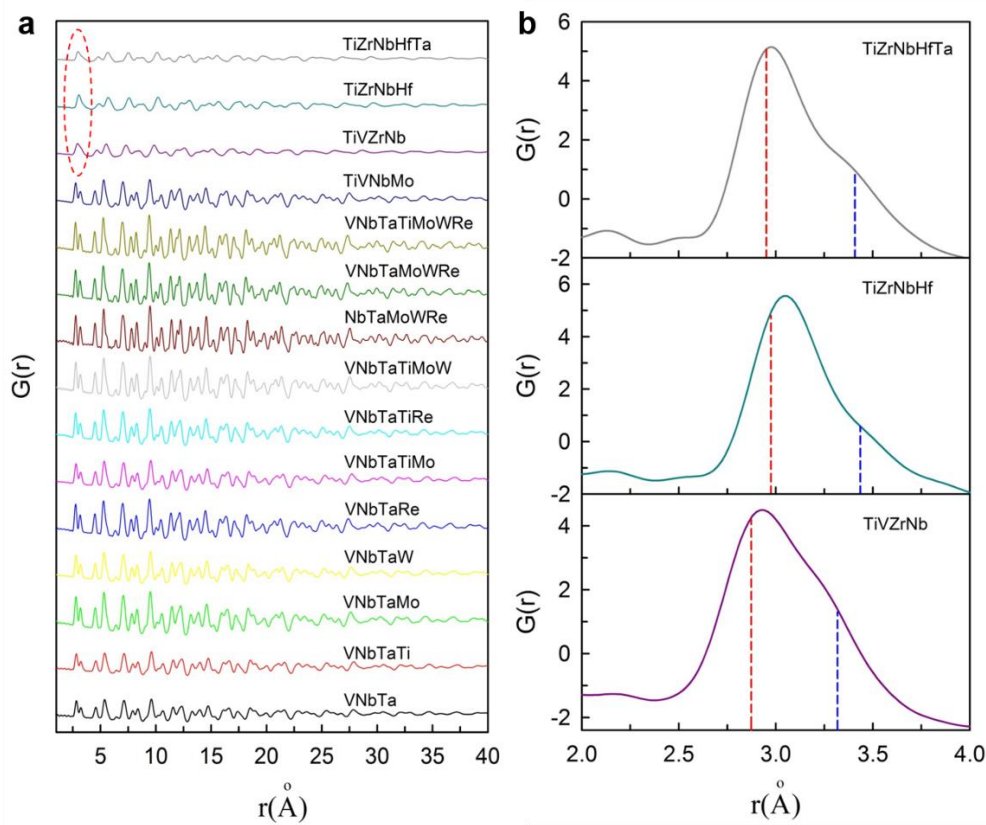

**Figure 2. Local structure of the RCSAs.** (a) PDFs of the RCSAs. The dash-line ellipse highlights the overlapped first and second atomic shells in the ZH-RCSAs. (b) The first and second atomic shells in the TiZrNbHfTa, TiZrNbHf and TiVZrNb RCSAs. The red and blue dash lines indicate the ideal positions of the first and second shells, respectively, calculated from the lattice constant.

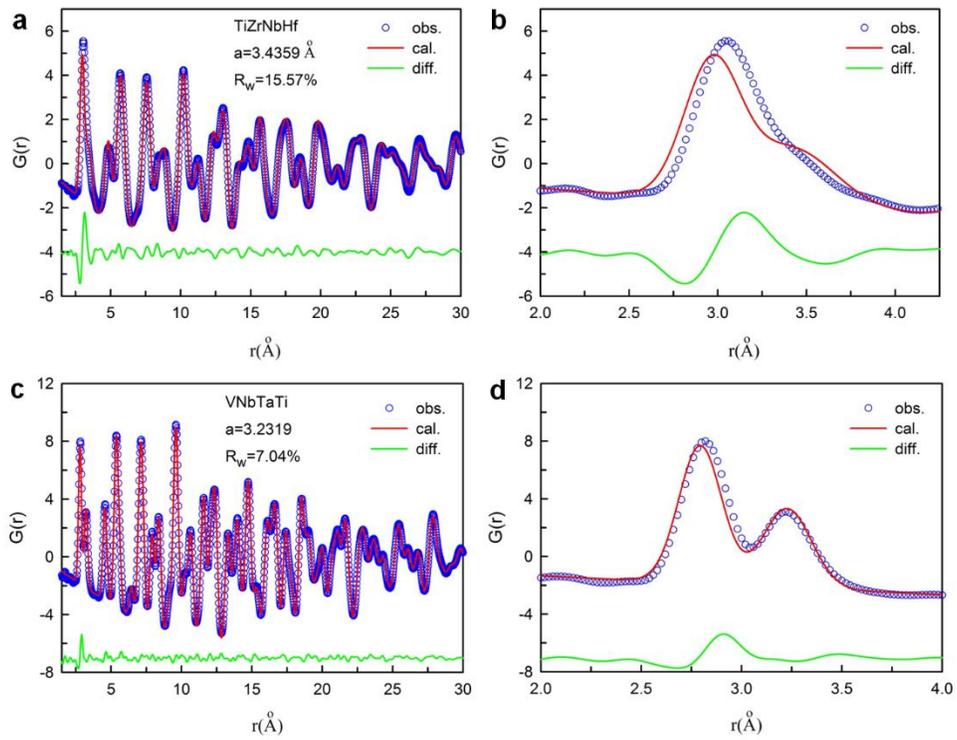

**Figure 3. Deviation of local structure from average structure.** A comparison of the measured PDFs and their fits based on undistorted BCC random-alloy structure for TiZrNbHf (a) and VNbTaTi (c) RCSAs. (b) and (d) shows an enlarged view of the local structure deviation in the TiZrNbHf and VNbTaTi RCSAs, respectively.

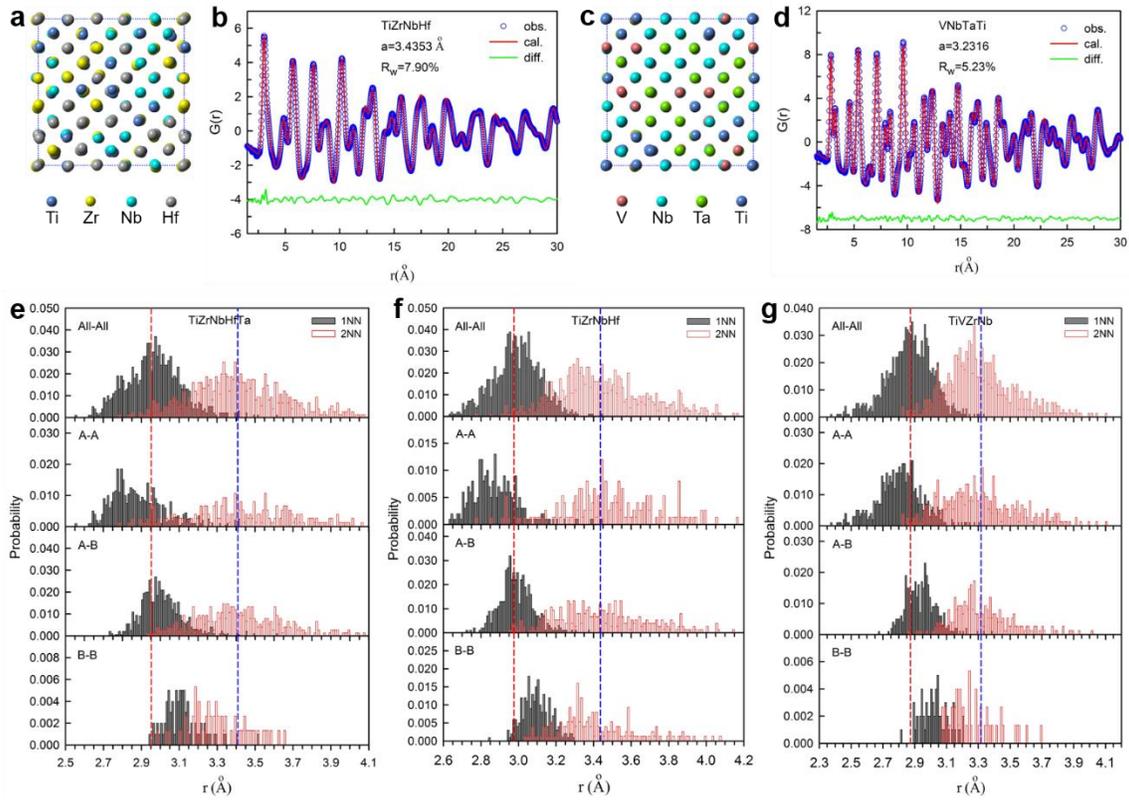

**Figure 4. Distorted structure model.** A comparison of the measured PDFs and their fits based on distorted BCC random-alloy structure (a, c) for TiZrNbHf (b) and TiVNbTa (d). (e, f, g) Distance distribution of total (All-All) and partial (A-A, A-B and B-B) atomic pairs in the first and second atomic shells of TiZrNbHfTa, TiZrNbHf and TiVZrNb RCSAs. A represents Ti, Nb, Ta and/or V element and B denotes Zr and/or Hf. The red and blue dash lines indicate the ideal positions of the first and second shells, respectively, expected from the lattice constant of the average structure.

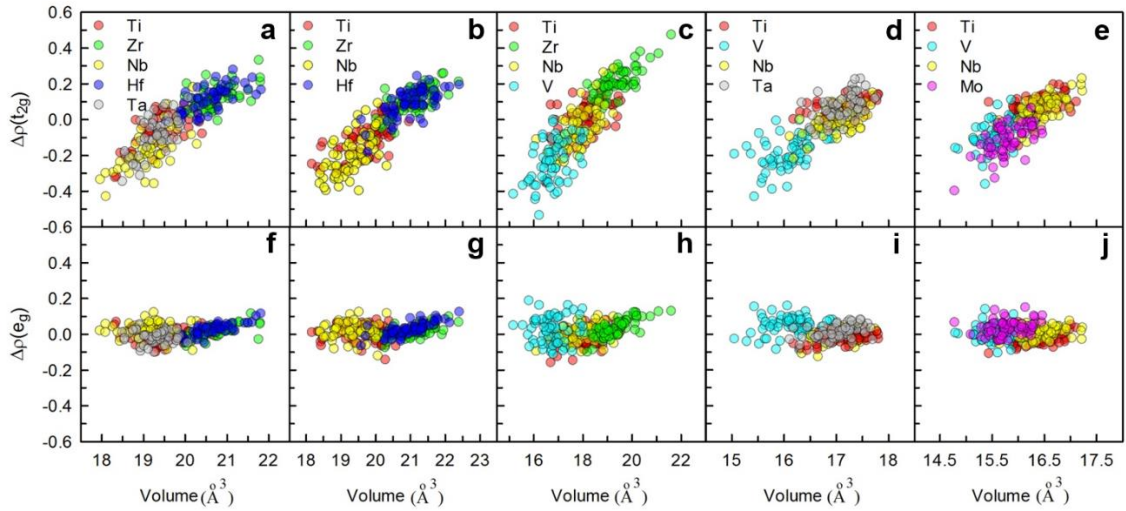

**Figure 5. Relation between charge transfer and Wigner-Seitz volume.** Correlations between atomic volumes and charge transfer of $t_{2g}$ (upper panel) and $e_g$ (lower panel) orbitals for TiZrNbHfTa (a,f), TiZrNbHf (b,g), TiZrNbV (c,h), TiVNbTa (d,i) and TiVNbMo (e,j) RCSAs.

**Table 1.** Structure and LLD parameters.

| Composition | Lattice constant | $U_{iso}$ (Å$^2$) | $\varepsilon_{1st}$ (%) |
|---|---|---|---|
| TiZrNbHfTa | 3.4088 | 0.0291 | 0.91 |
| TiZrNbHf | 3.4359 | 0.0292 | 2.39 |
| TiVZrNb | 3.3181 | 0.0302 | 2.00 |
| TiVNbMo | 3.1846 | 0.0144 | 0.01 |
| VNbTaTiMoWRe | 3.1678 | 0.0083 | 0.08 |
| VNbTaMoWRe | 3.1653 | 0.0077 | 0.23 |
| NbTaMoWRe | 3.1901 | 0.0064 | 0.05 |
| VNbTaTiMoW | 3.1911 | 0.0106 | 0.21 |
| VNbTaTiRe | 3.1865 | 0.0136 | 0.02 |
| VNbTaTiMo | 3.2119 | 0.0148 | 0.52 |
| VNbTaRe | 3.1750 | 0.0110 | 0.22 |
| VNbTaW | 3.2043 | 0.0126 | 0.32 |
| VNbTaMo | 3.1998 | 0.0126 | 0.65 |
| VNbTaTi | 3.2319 | 0.0150 | 0.82 |
| VNbTa | 3.2291 | 0.0146 | 0.99 |

Note: Lattice constant and thermal factor, $U_{iso}$, are obtained from the refinement of average structure. Local lattice distortion parameter, $\varepsilon_{1st}$, is from the refinement of local structure.

Table 2. WS radii of each element in BCC refractory metals and RCSAs.

| Composition | WS radius (Å) | | | | | | |
|---|---|---|---|---|---|---|---|
| | Ti | V | Zr | Nb | Mo | Hf | Ta |
| Ti | 1.63 | — | — | — | — | — | — |
| V | — | 1.49 | — | — | — | — | — |
| Zr | — | — | 1.78 | — | — | — | — |
| Nb | — | — | — | 1.63 | — | — | — |
| Mo | — | — | — | — | 1.55 | — | — |
| Hf | — | — | — | — | — | 1.78 | — |
| Ta | — | — | — | — | — | — | 1.63 |
| TiZrNbHfTa | 1.67 | — | 1.70 | 1.66 | — | 1.70 | 1.66 |
| TiZrNbHf | 1.67 | — | 1.71 | 1.66 | — | 1.71 | — |
| TiVZrNb | 1.63 | 1.59 | 1.67 | 1.62 | — | — | — |
| TiVNbTa | 1.60 | 1.57 | — | 1.60 | — | — | 1.60 |
| TiVNbMo | 1.57 | 1.55 | — | 1.58 | 1.56 | — | — |

Note: WS radius of pure metals is calculated from their lattice constants.